\documentclass[
  journal=nalefd,
  manuscript=letter
  ]{achemso}
\setkeys{acs}{doi = true}

\usepackage{chemformula} 
\usepackage[T1]{fontenc} 
\usepackage[colorlinks,citecolor=magenta]{hyperref}

\author{Chenjiang Qian}
\email{chenjiang.qian@wsi.tum.de}
\affiliation{Walter Schottky Institut and Physik Department, Technische Universit{\"a}t M{\"u}nchen, Am Coulombwall 4, 85748 Garching, Germany}
\author{Viviana Villafañe}
\affiliation{Walter Schottky Institut and Physik Department, Technische Universit{\"a}t M{\"u}nchen, Am Coulombwall 4, 85748 Garching, Germany}
\author{Martin Schalk}
\affiliation{Walter Schottky Institut and Physik Department, Technische Universit{\"a}t M{\"u}nchen, Am Coulombwall 4, 85748 Garching, Germany}
\author{G. V. Astakhov}
\affiliation{Helmholtz-Zentrum Dresden-Rossendorf, Institute of Ion Beam Physics and Materials Research, 01328 Dresden, Germany}
\author{Ulrich Kentsch}
\affiliation{Helmholtz-Zentrum Dresden-Rossendorf, Institute of Ion Beam Physics and Materials Research, 01328 Dresden, Germany}
\author{Manfred Helm}
\affiliation{Helmholtz-Zentrum Dresden-Rossendorf, Institute of Ion Beam Physics and Materials Research, 01328 Dresden, Germany}
\author{Pedro Soubelet}
\affiliation{Walter Schottky Institut and Physik Department, Technische Universit{\"a}t M{\"u}nchen, Am Coulombwall 4, 85748 Garching, Germany}
\author{Nathan P. Wilson}
\affiliation{Walter Schottky Institut and Physik Department, Technische Universit{\"a}t M{\"u}nchen, Am Coulombwall 4, 85748 Garching, Germany}
\author{Roberto Rizzato}
\affiliation{Department of Chemistry, Technical University of Munich, Lichtenbergstr. 4, Garching 85748, Germany}
\author{Stephan Mohr}
\affiliation{Department of Chemistry, Technical University of Munich, Lichtenbergstr. 4, Garching 85748, Germany}
\author{Alexander W. Holleitner}
\affiliation{Walter Schottky Institut and Physik Department, Technische Universit{\"a}t M{\"u}nchen, Am Coulombwall 4, 85748 Garching, Germany}
\author{Dominik B. Bucher}
\affiliation{Department of Chemistry, Technical University of Munich, Lichtenbergstr. 4, Garching 85748, Germany}
\author{Andreas V. Stier}
\affiliation{Walter Schottky Institut and Physik Department, Technische Universit{\"a}t M{\"u}nchen, Am Coulombwall 4, 85748 Garching, Germany}
\author{Jonathan J. Finley}
\email{finley@wsi.tum.de}
\affiliation{Walter Schottky Institut and Physik Department, Technische Universit{\"a}t M{\"u}nchen, Am Coulombwall 4, 85748 Garching, Germany}

\title {Unveiling the Zero-Phonon Line of the Boron Vacancy Center by Cavity Enhanced Emission}


\begin{document}

\begin{tocentry}

  \includegraphics[width=\linewidth]{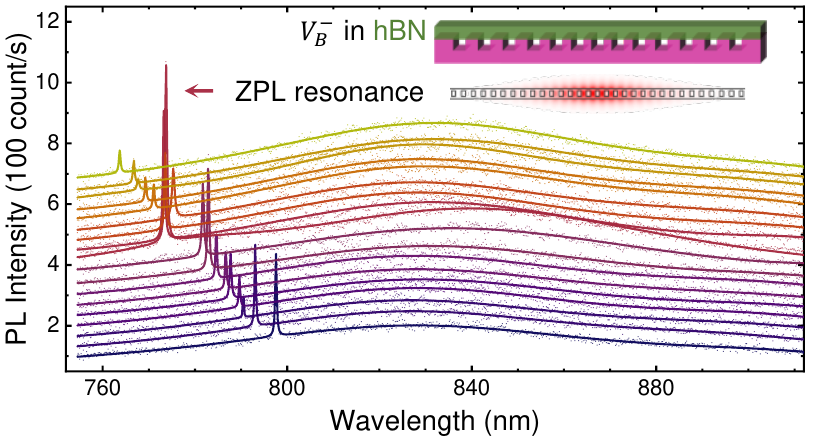}

\end{tocentry}

\begin{abstract}
  Negatively charged boron vacancies ($V_B^-$) in hexagonal boron nitride (hBN) exhibit a broad emission spectrum due to strong electron-phonon coupling and Jahn-Teller mixing of electronic states.
  As such, the direct measurement of zero-phonon line (ZPL) of $V_B^-$ has remained elusive.
  Here, we measure the room-temperature ZPL wavelength to be $773\pm2$ nm by coupling the hBN layer to the high-Q nanobeam cavity.
  As the wavelength of cavity mode is tuned, we observe a pronounced intensity resonance, indicating the coupling to $V_B^-$.
  Our observations are consistent with the spatial redistribution of $V_B^-$ emission.
  Spatially resolved measurements show a clear Purcell effect maximum at the midpoint of the nanobeam, in accord with the optical field distribution of the cavity mode.
  Our results are in good agreement with theoretical calculations, opening the way to using $V_B^-$ as cavity spin-photon interfaces.

\end{abstract}

\section{Introduction}

Negatively charged boron vacancies ($V_B^-$) in hexagonal boron nitride (hBN) are novel quantum emitters that currently attract broad interest due to their spin-dependent optical properties, even at room temperature \cite{Gottscholl2020,Gottscholl2021,doi:10.1021/acsphotonics.1c00320,doi:10.1021/acsomega.1c04564,2201.13184}.
The electronic states of $V_B^-$ in hBN couple strongly to local vibrational modes \cite{Ivady2020,PhysRevB.102.144105,PhysRevLett.128.167401}.
Combined with singlet and triplet electronic sub-systems that are coupled via spin-orbit interaction and mix via Jahn-Teller effects, this results in a broad and largely featureless emission spectrum that complicates comparison of experimental results with theoretical calculations \cite{Ivady2020,doi:10.1021/acs.jpca.0c07339,PhysRevB.102.144105,PhysRevLett.128.167401,2103.16494}.
Typically, the photoluminescence (PL) spectrum of ensembles of $V_B^-$ exhibit a broad and featureless spectrum dominated by phonon-assisted emissions peaked around 800 nm \cite{doi:10.1021/acsomega.1c04564,2201.13184,Ivady2020,PhysRevB.102.144105,PhysRevLett.128.167401} even at liquid helium temperature \cite{2004.07968}.
Therefore, it is difficult to unambiguously identify the zero-phonon line (ZPL) of $V_B^-$ from transitions involving vibrational modes of the surrounding hBN matrix, in contrast to the nitrogen vacancies where the ZPL peak can be directly distinguished in the PL spectra \cite{PhysRevApplied.5.034005}.

In this letter, we unambiguously identify the ZPL of $V_B^-$ by coupling micro ensembles to the cavity mode of a proximal nanobeam cavity.
The presence of $V_B^-$ in our hBN samples is confirmed by performing optically detected magnetic resonance (ODMR) and photoluminescence spectroscopy.
We exploit cavity QED phenomena to trace out the ZPL by recording detuning-dependent PL spectra, where the cavity-ZPL detuning is controlled via the geometric width of the cavity ($d_y$).
We observe a pronounced maximum of the intensity of cavity luminescence at $773\pm2$ nm with a strongly asymmetric lineshape and linewidth $\leq10$ nm.
This observation is indicative of resonant enhancement arising due to cavity QED phenomena induced by the proximal cavity.
The experimentally observed ZPL wavelength agrees well with values in previous works, which are obtained by fitting the overall broad $V_B^-$ emission with various theoretical models \cite{Ivady2020,PhysRevB.102.144105,PhysRevLett.128.167401}.
We further explore the nature of the cavity-ZPL coupling by measuring position-dependent PL spectra along a single nanobeam cavity.
We observe that the Purcell enhancement of the ZPL emission quantitatively agrees with the optical field distribution of the cavity \cite{PhysRevB.60.13276,Lee2015,Vukovi2017}.
In contrast, when the cavity mode is detuned from the ZPL into the phonon sideband (PSB), no similar resonant intensity enhancement is observed.
This observation is shown to arise from the short lifetime polaritonic states in the PSB that results in much weaker coupling to the cavity mode \cite{PhysRevB.65.235311,PhysRevB.60.13276,Lee2015,Vukovi2017}.
Similarly weak cavity-PSB coupling (as compared to cavity-ZPL) has been reported for nitrogen vacancies in hBN \cite{doi:10.1063/5.0046080} and color centers in diamond \cite{Faraon2011,Johnson_2015}.
Therefore, the coupling to the cavity mode enhances and distinguishes the ZPL of $V_B^-$, which is a fundamental property for quantum emitters.
Moreover, our clear observed cavity-ZPL coupling enhances the fraction of energy directed into the ZPL, making luminescent atomic scale defects such as $V_B^-$ interesting as prototypical, room temperature spin-photon interfaces \cite{PhysRevLett.95.030504,PhysRevLett.105.033601}.

\section{Results and discussion}

\begin{figure}
  \includegraphics[width=0.8\linewidth]{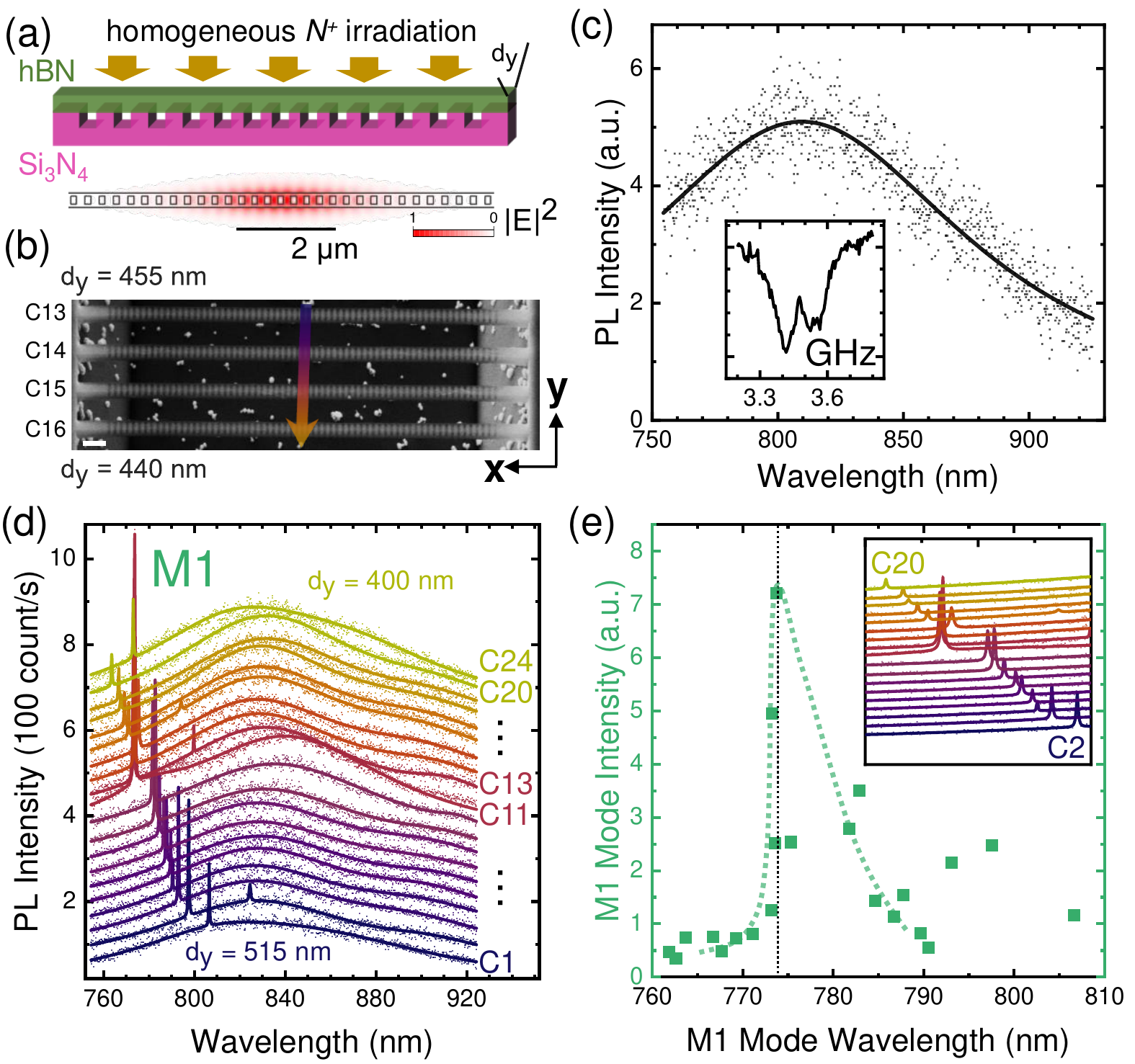}
  \caption{
    (a) Schematic of the cavity and the electric field $\vert E\vert^2$ distribution of the cavity mode M1 around the cavity center.
    All trenches have the width of 140 nm and depth of 120 nm.
    The cavity mode volume along the nanobeam is $2\ \mathrm{\mu m}$.
    The hBN flake is homogeneously irradiated by nitrogen ions.
    (b) SEM image of cavities with varying width $d_y$.
    White scale bar is $1\ \mathrm{\mu m}$.
    The occurrence of $V_B^-$ is identified by the PL and ODMR spectra in (c).
    The arrow in (b) marks the detuning-dependent PL spectra measured at the center of each cavity, and the results are plotted in (d).
    (e) The cavity intensity vs. the mode wavelength extracted by multi-Lorentz fitting.
    The sharp peak of cavity intensity around 773 nm reveals the resonance to ZPL emission of $V_B^-$.
  }
  \label{f1}
\end{figure}

The hBN flakes used in our experiments had a thickness of 100 nm and were transferred from natural crystals using standard viscoelastic dry transfer methods.
The structure of our hybrid nanobeam cavities and the calculated cavity field distribution of the first mode M1 are presented in Fig.~\ref{f1}(a).
The nanobeam cavities were prepared by firstly etching corrugations into the 200 nm thick Si$_3$N$_4$ to form a one dimensional grating, before the hBN layer was transferred on top\cite{PhysRevLett.128.237403}.
The nanobeam resonators were then defined via electron beam lithography and etching, before the underlying silicon was removed using wet etching.
The nanobeam width ($d_{y}$) was varied between $400-515$ nm in the array of cavities to precisely tune the wavelength of the cavity mode. SEM images of several nanobeam cavities are presented in Fig.~\ref{f1}(b). Further details of the fabrication processes and the cavity mode properties can be found in ref \cite{PhysRevLett.128.237403}.
After the transfer, the hBN was homogeneously irradiated by nitrogen ions with the energy of 30 keV and the fluence dose of $10^{14}\ \mathrm{ions/\mu m^2}$.
We confirmed the presence of $V_B^-$ defects in our samples by performing room temperature PL and optically detected magnetic resonance (ODMR) spectroscopy on the irradiated hBN crystal.
Typical results are presented in Fig.~\ref{f1}(c) that shows the typical broad photoluminescence recorded at room temperature with a maximum around 800 nm.
The inset of Fig.~\ref{f1}(c) shows a typical continuous wave ODMR spectrum exhibiting two characteristic transitions in a zero magnetic field, arising from redistribution of spin amongst the triplet $m_S=0$ and $m_s=\pm1$ triplet ground state manifold of $V_B^-$ \cite{Ivady2020,2201.13184,Gottscholl2020,Gottscholl2021,doi:10.1021/acsphotonics.1c00320,doi:10.1021/acsomega.1c04564}.

Figure~\ref{f1}(d) shows the PL spectra of the successive cavities marked C1-C24.
Cavities C12,21-23 were broken during the fabrication and thus not considered here.
As depicted schematically by the arrow on Fig.~\ref{f1}(b), each spectrum was recorded at the center of the cavity where the maximum intensity of the cavity mode M1 is observed.
Although the $V_B^-$ defect is usually excited with laser power densities in the 100 kW/cm$^2$ range or higher \cite{2004.07968,doi:10.1021/acsomega.1c04564}, here we excite $V_B^-$ with a 532-nm laser and a power density of 35 kW/cm$^2$ to suppress non-linear effects and laser-induced heating/oxidation as much as possible.
As expected, we clearly observe a systematic blueshift of the cavity mode M1 with decreasing $d_y$ (Fig.~\ref{f1}(d)).
For some cavities, e.g. C2 and C11, another minor cavity mode M2 at the red side of M1 is observed.
M2 has a low Q $\sim$ one order of magnitude smaller than M1 and a zero electric field at the cavity center \cite{PhysRevLett.128.237403}, thereby its coupling to the $V_B^-$ defect is weak and out of the scope of this work.

As the cavity mode frequency is tuned for nominally identical experimental conditions its intensity exhibits a pronounced maximum.
In the data presented in Fig.~\ref{f1}(d) and (e), this occurs for cavity C11.
The maximum intensity occurs for $\lambda_{M1}=773$ nm, as presented in Fig.~\ref{f1}(e).
The observation of a resonance in the intensity of the cavity mode is indicative of a spatial redistribution of the radiation from a specific dipole allowed transition in the hBN arising, for example, due to the Purcell effect.
Ivády et al. \cite{Ivady2020} reported the ZPL of $V_B^-$ to be at 765 nm by aligning calculated ${^3E''}\rightarrow {^3A_2'}$ triplet emission obtained from the ab inito theory. By using a Hyang-Rhys factor of 3.5 the data was fitted to a measured PL spectrum to obtain good agreement on the high-energy side of the spectrum.
Reimers et al. \cite{PhysRevB.102.144105} reported the ZPL to be at 770 nm by aligning calculated ${^3E''}\rightarrow {^3A_2'}$ triplet emission based on the Huang-Rhys model, and achieved good agreement on the low-energy side of the spectrum.
Libbi et al. \cite{PhysRevLett.128.167401} calculated the $V_B^-$ emission using many-body perturbation theory, and achieved good agreement with experiment by aligning the ZPL to 756 nm.
Each of these reported wavelengths for the ZPL are close to the observed resonance in our experiment.
As such, we ascribe the observed resonance in the M1 mode intensity as arising from resonant coupling with the ZPL of $V_B^-$.
Since the $V_B^-$ defects are localized emitters, their emission can be modeled using the dipole approximation \cite{PhysRevB.86.085304,PhysRevLett.122.087401}. In this picture, the cavity-emitter coupling is proportional to the cavity quality factor (Q-factor), the square of the interaction strength $\vert\mathbf{d}\cdot\alpha\left(\mathbf{r}\right)\vert^2$, and an additional Lorentzian factor accounting for the cavity-emitter detuning \cite{PhysRevB.60.13276,Lee2015,Vukovi2017}.
Here $\mathbf{d}$ is the constant intrinsic emission dipole moment of $V_B^-$, and $\alpha\left(\mathbf{r}\right)$ is the cavity mode function proportional to the cavity electric field at the emitter position $\mathbf{r}$, as depicted in Fig.~\ref{f1}(a).
Both the amplitude of the cavity vacuum field ($\alpha_{max}$) and the Q-factor vary only weakly ($<30\%$) across all cavities examined.
Therefore, we conclude that the observed strongly resonant behavior of the cavity intensity (Fig.~\ref{f1}(e)) must originate from the cavity-emitter detuning between the ZPL and the cavity mode.
The observed detuning-dependence resonance indicates that the ZPL is centered $773\pm2$ nm.
In addition, when the cavity mode is red-detuned from the ZPL ($>785$ nm), it couples to the phonon sideband.
However, the short lifetime polaritonic states in the PSB \cite{PhysRevB.65.235311} results in much weaker coupling to the cavity mode \cite{PhysRevB.60.13276,Lee2015,Vukovi2017}.
This is similar to the dominant role of the cavity-ZPL coupling that has been reported for the nitrogen vacancies in hBN \cite{doi:10.1063/5.0046080} and the nitrogen-vacancy center in diamond \cite{Faraon2011,Johnson_2015}.
As a result, although the PSB emission dominates the broad PL spectra in Fig.~\ref{f1}(d), the intensity of the cavity modes in the PSB emission is comparatively weak.

\begin{figure}
  \includegraphics[width=0.8\linewidth]{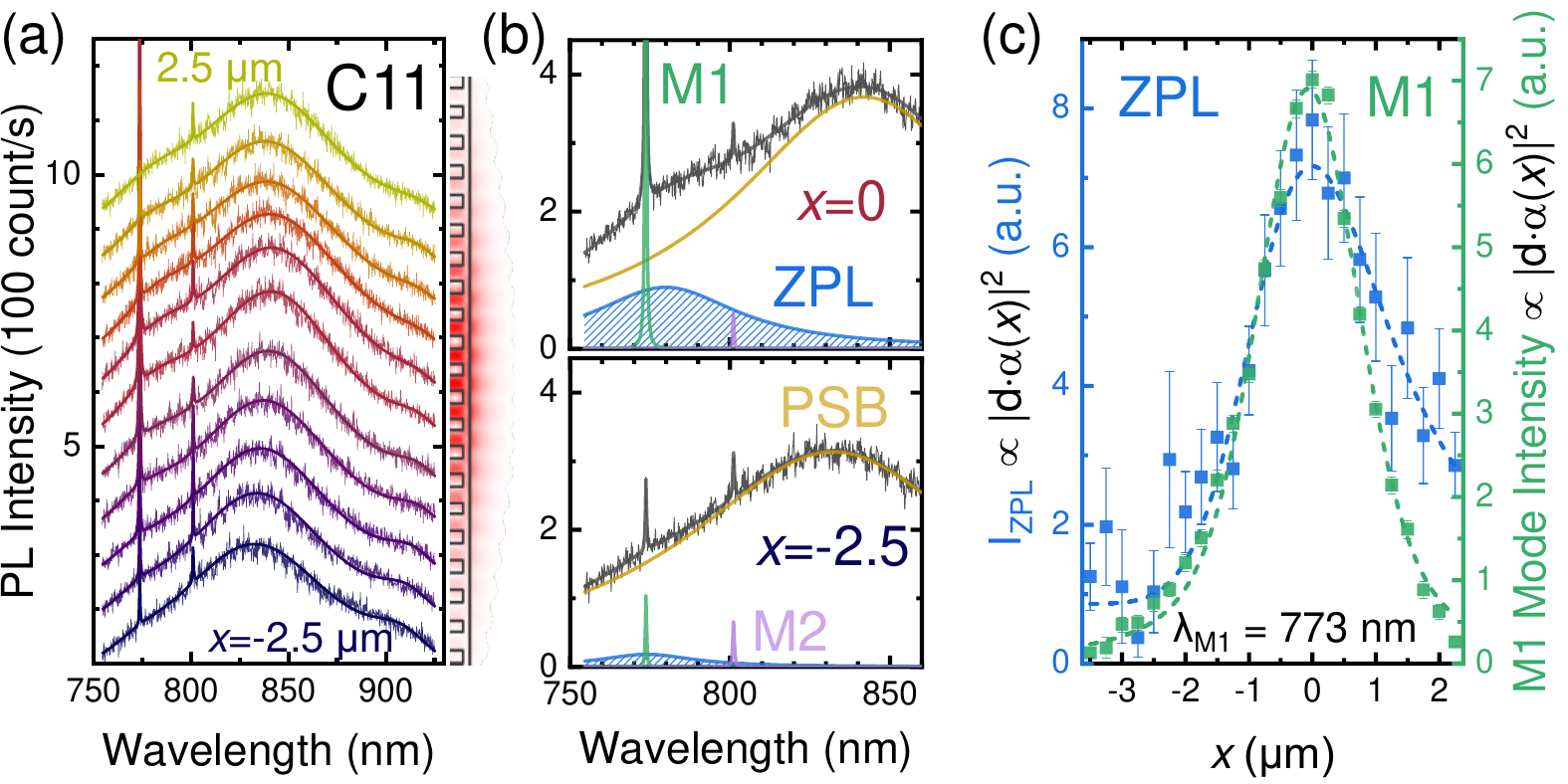}
  \caption{
    Purcell enhancement of the ZPL observed by scanning the laser spot along the nanobeam cavity, throughout the spatial profile of the cavity mode.
    (a) Position-dependent PL spectra in cavity C11.
    The corresponding cavity mode function $\vert\alpha\left(x\right)\vert^2$ (proportional to the electric field in Fig.~\ref{f1}(a)) is plotted at rightmost edge of the panel.
    An additional peak appears at the cavity center.
    (b) PL spectra and multi-Lorentz fittings at center ($x=0\ \mathrm{\mu m}$) and away from the center ($x=-2.5\ \mathrm{\mu m}$) for better comparison.
    The blue shaded peak is the ZPL and green peak is the M1 cavity mode.
    The Purcell enhancement of the ZPL peak at the center of the nanobeam is clearly observed.
    (c) Spatial dependence of the intensity of the ZPL (blue) and cavity mode M1 (green) for cavity C11. Both features exhibit same spatial dependence demonstrating that the ZPL intensity is proportional to $\vert\alpha\left(x\right)\vert^2$ in our experiments.
  }
  \label{f2}
\end{figure}

The conclusion of cavity-ZPL coupling is further supported by the position-dependent PL spectra as the excitation laser spot is scanned along the length of a nanobeam.
Typical data is presented in Fig.~\ref{f2}(a).
The intensity of the cavity mode M1 exhibits a pronounced maximum at the mid-point of the nanobeam (Fig.~\ref{f2}(b)), revealing the maximum cavity mode function $\alpha\left(x\right)$ (Fig.~\ref{f1}(a)) at the center \cite{PhysRevLett.128.237403}.
In addition to the enhanced cavity mode peak, for cavity C11 ($\lambda_{M1}=773\ \mathrm{nm}$) we observe an additional PL peak shown in the expanded spectra with multi-Lorentz fittings in Fig.~\ref{f2}(a)(b).
The Purcell enhancement of the additional peak (blue) is clearly observed, revealing that this peak is indeed the ZPL peak broadened by the continuum of low-frequency phonons and the ensemble of emitters.
This is the central result of this Letter.
The cavity-enhanced emission rate is given by
\begin{equation}
  \gamma_{SE}=\frac{8\pi Q}{\hbar}\frac{\vert\mathbf{d}\cdot\alpha\left(x\right)\vert^2}{4\pi \epsilon_r\epsilon_0}
  \label{eq1}
\end{equation}
where $\epsilon_r$ ($\epsilon_0$) is the relative (vacuum) permittivity.
As presented in Fig.~\ref{f2}(c), the ZPL intensity very closely follows the cavity-emitter coupling strength $\vert\mathbf{d}\cdot\alpha\left(x\right)\vert^2$ (equivalent to the cavity intensity), exhibiting a $\approx 21\mathrm{x}$ enhancement at the cavity center for which the cavity field $\alpha\left(x\right)$ is maximum.

\begin{figure}
  \includegraphics[width=0.8\linewidth]{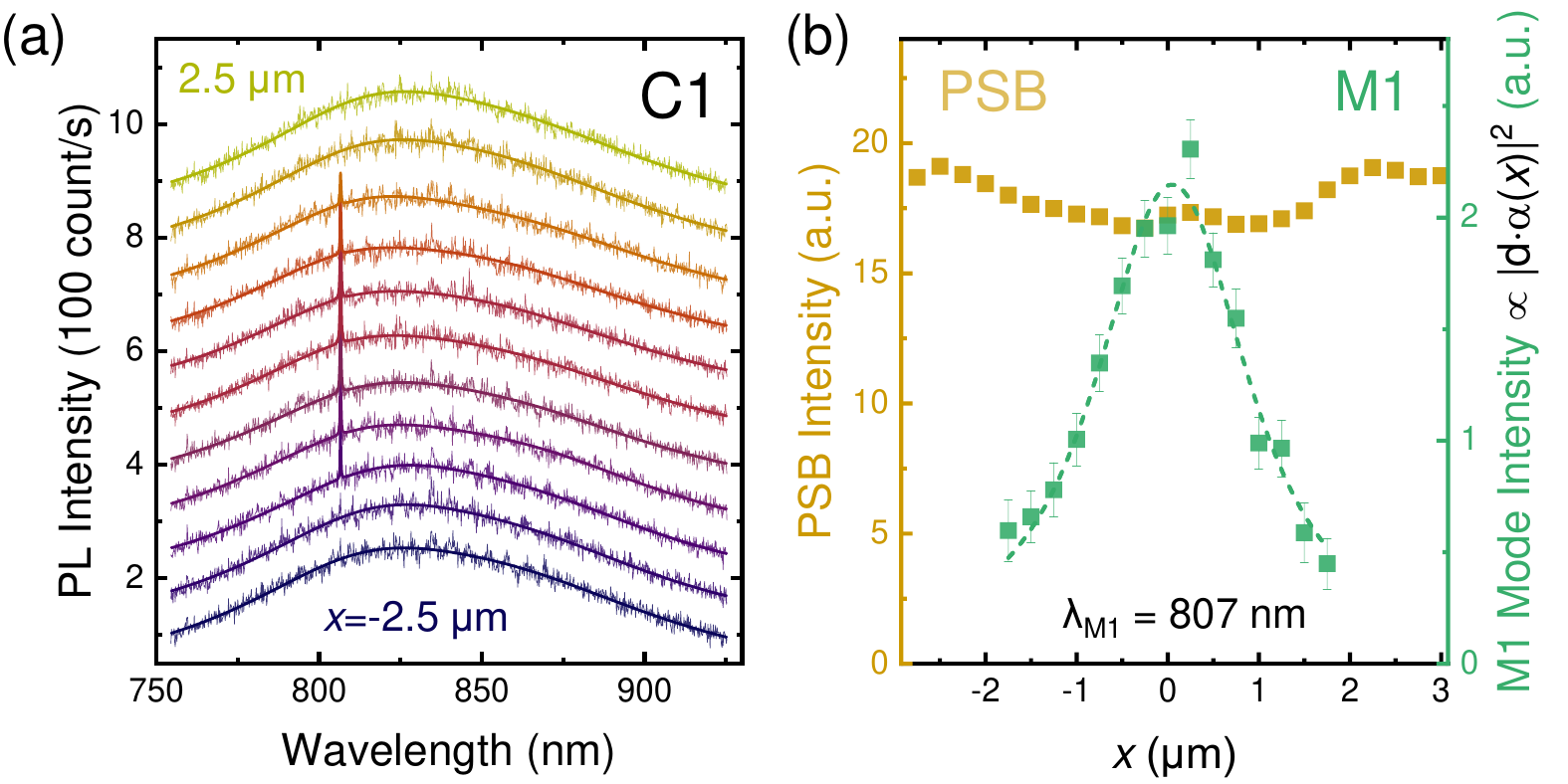}
  \caption{
    Demonstration that an intensity enhancement does not occur for the PSB as the laser spot is spatially scanned along the nanobeam, over the spatial profile of the cavity mode.
    (a) Position-dependent PL spectra recorded from cavity C1.
    (b) Spatial variation of the intensity of the PSB (yellow) and cavity mode M1 (green) in C1 as the laser is scanned along the nanobeam. Little enhancement of the PSB emission (yellow) is observed.
    Here, we use the PL intensity due to PSB emissions dominate the PL spectra, and the sum intensity was found to be more accurate than multi-Lorentz fitting of broad peaks.
  }
  \label{f3}
\end{figure}

In contrast, when the cavity mode is tuned into the broad PSB emission, such as in the case for cavity C1 ($\lambda_{M1}=807\ \mathrm{nm}$), the position-dependent resonance in the intensity of the broad PSB emission is not observed (see Fig.~\ref{f3}(a)(b)).
This observation is fully consistent with the weak cavity-PSB coupling, as discussed in connection with Fig.~\ref{f1}.
Indeed, the PSB emission exhibits a minor suppression as shown in Fig.~\ref{f3}(b), similar to observations for nitrogen-vacancy centers in diamond as has been reported previously arsing from the suppression of the optical density of states \cite{Johnson_2015}.

\begin{figure}
  \includegraphics[width=0.8\linewidth]{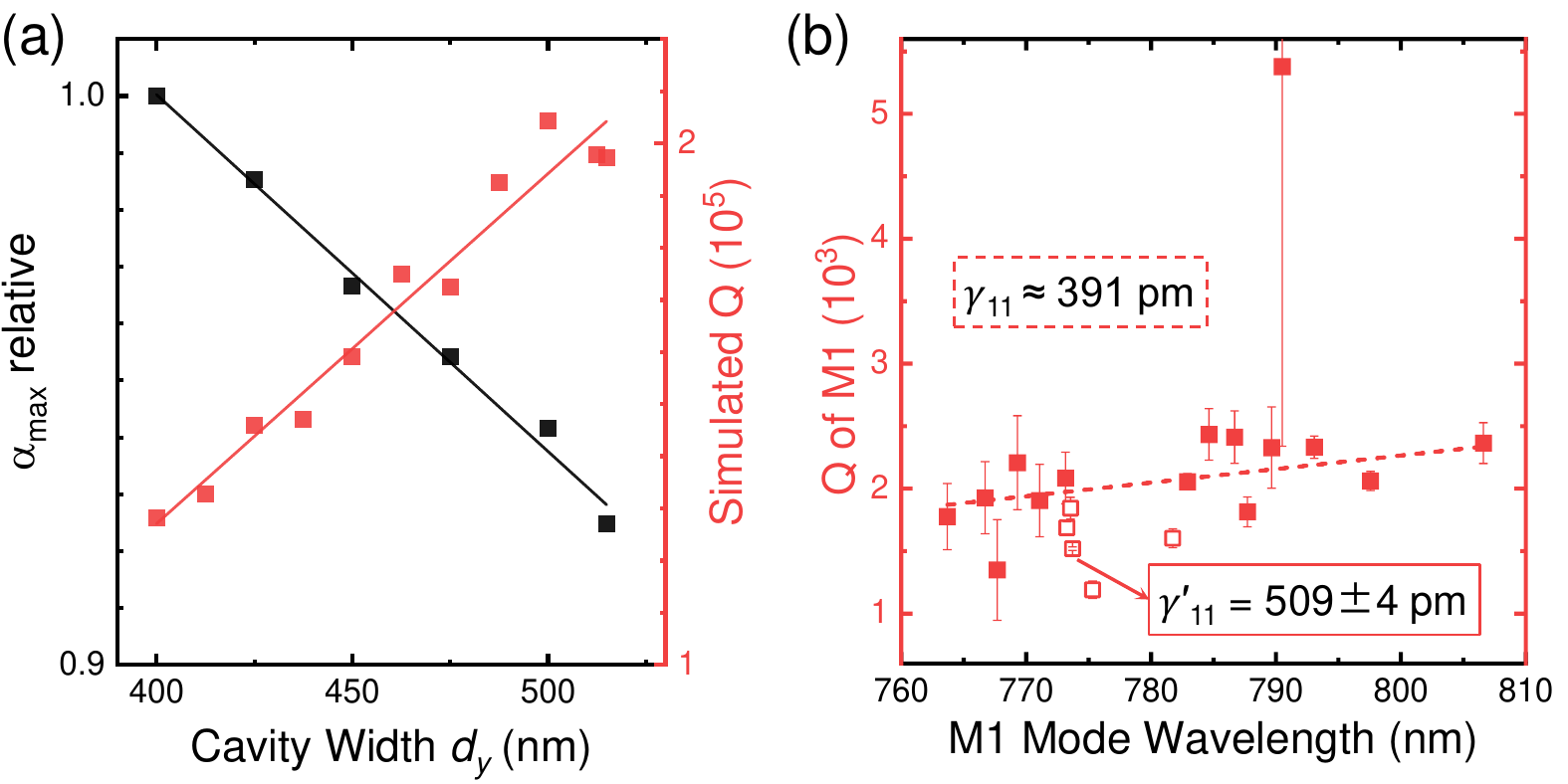}
  \caption{
    (a) Simulated cavity mode function and Q-value by the FDTD method, exhibiting a near linear variation with the cavity width $d_y$.
    Here $\alpha_{max}$ is plot with the ratio to the value at $d_y=400\ \mathrm{nm}$ for better comparison.
    (b) Estimation of the bare cavity linewidth of C11 by the linear fitting (dashed line) of Q-values in far-detuned cavities (solid dots).
    The Q-values at hollow dots have less accuracy due to the the cavity linewidth broadening near the cavity-ZPL resonance.
  }
  \label{f4}
\end{figure}

The detuning and position dependent PL data presented in Figs.~\ref{f1} and \ref{f2} provide strong evidence that the local vacuum field enhancement in the cavity is responsible for the observed resonant behaviour of the ZPL intensity.
We continue to estimate the coupling strength $g$ between the ZPL and cavity mode by tracking the linewidth broadening expected from the Jaynes–Cummings model of cavity QED.
Hereby, the linewidth of the polariton mode is given by $\gamma_{11}'=\gamma_{11}+4g^2/\gamma_X$ at resonance (cavity C11) where $\gamma_{11}'=509 \pm 4\ \mathrm{pm}$ (extracted by fitting) is the cavity linewidth after coupling, $\gamma_{11}=\lambda_{11}/Q_{11}$ is the uncoupled cavity linewidth and $\gamma_X$ is the intrinsic linewidth of the emitter.
We performed FDTD simulations to calculate the expected dependence of $\gamma_{11}$ on $d_y$. Typical results of our simulations are presented in Fig.~\ref{f4}(a) that reveals a weak linear dependence of the mode Q-factor on $d_y$.  Since the variation of $d_y$ in our experiments is small ($<20\%$), the cavity Q and the maximum cavity field $\alpha_{max}$ are both expected to be linear functions of $d_y$ \cite{PhysRevLett.128.237403}.
Thus, as described by the dashed line in Fig.~\ref{f4}(b) we estimated $\gamma_{11}$ by fitting the cavity Q-values in far-detuned cavities (solid dots), and extrapolating the uncoupled cavity linewidth for the cavity on resonance with the ZPL.
The procedure gives the result $\gamma_{11}=391\ \mathrm{pm}$.
The emitter linewidth $\gamma_X$ is estimated from the resonant peak in Fig.~\ref{f1}(e).
We note that, the highly asymmetric lineshape in Fig.~\ref{f1}(e) most likely originates from coupling to a continuum of acoustic phonons.  The spectral width of the acoustic phonon sideband is related to the realspace extent of the involve electronic wavefunction, as observed for atomic scale emitters in other 2D materials, including MoS$_2$ monolayers \cite{Klein2019,doi:10.1021/acsphotonics.0c01907}.
Here for brevity, we take the FWHM of the whole peak (8 nm) as a maximum and the FWHM at the sharp blue side (1 nm) as a minimum to estimate $\gamma_X$ in the Jaynes–Cummings model.
With the estimated values of $\gamma_{11}$ and $\gamma_X$, the interaction strength $g$ is calculated to be $0.4-1$ meV ($0.17-0.48$ nm).
The density of boron vacancies generated by ions was estimated using SRIM simulations \cite{ARADI2013214} to be $\sim0.16$ vacancies per incident ion.
By comparing the dose of N-ions used to implant the hBN with the effective area of the cavity mode ($2\times0.5\ \mathrm{\mu m}^2$ in Fig.~\ref{f1}(a)), we obtain the number of emitters coupled to the cavity mode to be $N=1.6\times10^5$.
Thereby, we get the interaction strength of each emitter $g/\sqrt{N}=1-2.5\ \mathrm{\mu eV}$, which has the same magnitude compared to other defect color centers \cite{McCutcheon:08} and one or two order of magnitude smaller compared to traditional quantum dots \cite{RevModPhys.87.347,PhysRevLett.120.213901,PhysRevLett.122.087401}.
The small interaction strength of $V_B^-$ defect is consistent with the fact that the associated dipole moment of $V_B^-$ (single atom defect) is much smaller than that of typical quantum dots (tens of nm) \cite{PhysRevLett.103.127401}.
In addition, the mode function $\alpha_{max}$ in Fig.~\ref{f4}(a) only changes by $8.2\%$, and the variation of Q-values in Fig.~\ref{f4}(b) is $26\%$ if excluding one noisy point.
Therefore, in contrast to such small varying of $\alpha_{max}$ and Q, the variation of cavity intensity in Fig.~\ref{f1}(e) ($\approx$ 21 times between maximum and minimum) surely originates from the cavity-ZPL detuning, further supporting the conclusions.

\section{Conclusion}

In summary, we separate and identify the ZPL and PSB of $V_B^-$ emission from the broad featureless PL peak by the different cavity-ZPL (stronger) and cavity-PSB (weaker) coupling.
Detuning-dependent and position-dependent experimental results agree well to this difference and reveal the cavity-ZPL resonance at around 773 nm, a value close to the results in previous calculations \cite{Ivady2020,doi:10.1021/acs.jpca.0c07339,PhysRevB.102.144105,PhysRevLett.128.167401}.
This result indicate that the $V_B^-$ emission is dominated by the ${^3E''}\rightarrow {^3A_2'}$ transition with symmetry D3h \cite{Ivady2020,PhysRevB.102.144105}.
In addition, other minor effects are also observed, e.g. the PSB maximum exhibits some energy shift when the cavity-ZPL is near resonant.
These experimental results indicate the complex coupling between emitters, photons and phonons in the hBN-based cavity quantum electrodynamics systems \cite{PhysRevLett.126.227401} as a potential topic in future investigations.

\begin{acknowledgement}

  All authors gratefully acknowledge the German Science Foundation (DFG) for financial support via grants FI 947/8-1, DI 2013/5-1 and SPP-2244, as well as the clusters of excellence MCQST (EXS-2111) and e-conversion (EXS-2089).
  A. W. Holleitner and J. J. Finley acknowledge the state of Bavaria via the One Munich Strategy and Munich Quantum Valley.
  C. Qian and V. Villafañe gratefully acknowledge the Alexander v. Humboldt foundation for financial support in the framework of their fellowship programme.
  Support by the Ion Beam Center (IBC) at HZDR is gratefully acknowledged.

\end{acknowledgement}

\section{Measurement Setup}

Details about the design and fabrication of the hBN cavity can be found in ref. \cite{PhysRevLett.128.237403}.
The PL spectra of the hBN cavities are measured by a conventional confocal micro-PL setup.
The objective has a magnification of 100x and a NA of 0.75.
The sample is excited by a narrow linewidth solid state laser with the wavelength of 532 nm, the power of 0.35 mW and the laser spot size of $\sim1\ \mathrm{\mu m^2}$.
The sample position is aligned with a three-dimensional xyz nanopositioner with nm-scale resolution.
The PL spectra is collected with a matrix array Si CCD detector in a spectrometer with a focal length of 0.55 m and a grating of 300 grooves per mm.


\providecommand{\latin}[1]{#1}
\makeatletter
\providecommand{\doi}
  {\begingroup\let\do\@makeother\dospecials
  \catcode`\{=1 \catcode`\}=2 \doi@aux}
\providecommand{\doi@aux}[1]{\endgroup\texttt{#1}}
\makeatother
\providecommand*\mcitethebibliography{\thebibliography}
\csname @ifundefined\endcsname{endmcitethebibliography}
  {\let\endmcitethebibliography\endthebibliography}{}

\end{document}